\documentclass[12pt,a4paper]{article}
\usepackage[utf8]{inputenc}
\usepackage{lmodern}
\usepackage[T1]{fontenc}
\usepackage[english]{babel}
\usepackage{ifpdf}
\usepackage[usenames,dvipsnames]{xcolor}
\usepackage{graphicx}
\usepackage[shortlabels]{enumitem}
\usepackage[numbers,sort]{natbib}
\usepackage[font=footnotesize,width=\textwidth]{caption}
\usepackage{booktabs}
\usepackage{hyperref}
\usepackage{uri}
\usepackage{tikz}
\usepackage{float}
\usepackage[top=1.8cm, bottom=2.8cm]{geometry}
\usepackage[parfill]{parskip}
\usepackage[affil-it]{authblk}
\usepackage[setbuildercolon]{matmatix}
\newcommand{\G}{\mathcal G}
\newcommand{\T}{\mathcal T}

\ifpdf
  \hypersetup{
    pdftitle={xxx}
  }
\fi

\hypersetup{colorlinks, allcolors = {MidnightBlue}}

\makeatletter
\renewcommand*{\defas}{%
\mathrel{\rlap{\raisebox{0.3ex}{$\m@th\cdot$}}\raisebox{-0.3ex}{$\m@th\cdot$}}=}
\makeatother

\makeatletter
\newtheorem*{rep@theorem}{\rep@title}
\newcommand{\newreptheorem}[2]{%
\newenvironment{rep#1}[1]{%
 \def\rep@title{#2 \ref*{##1}}%
 \begin{rep@theorem}}%
 {\end{rep@theorem}}}
\makeatother

\makeatletter
\DeclareRobustCommand\bigop[2][1]{%
  \mathop{\vphantom{\sum}\mathpalette\bigop@{{#1}{#2}}}\slimits@
}
\newcommand{\bigop@}[2]{\bigop@@#1#2}
\newcommand{\bigop@@}[3]{%
  \vcenter{%
    \sbox\z@{$#1\sum$}%
    \hbox{\resizebox{\ifx#1\displaystyle#2\fi\dimexpr\ht\z@+\dp\z@}{!}{$\m@th#3$}}%
  }%
}
\makeatother

\newtheorem{theorem}{Theorem}
\newreptheorem{theorem}{Theorem}
\newtheorem{proposition}[theorem]{Proposition}
\newreptheorem{proposition}{Proposition}
\newtheorem{lemma}[theorem]{Lemma}
\newreptheorem{lemma}{Lemma}
\newtheorem{corollary}[theorem]{Corollary}
\theoremstyle{definition}

\newtheorem{remarkX}[theorem]{Remark}

\title{\vspace{-1cm}
{Asymptotic enumeration of normal and hybridization networks via tree decoration}}

\author [1] {Michael Fuchs}
\author[2]{Mike Steel}
\author[3]{Qiang Zhang}

\affil[1]{Department of Mathematical Sciences, National Chengchi University, Taipei 116, Taiwan}
\affil[2,3]{Biomathematics Research Centre, University of Canterbury, Christchurch, New~Zealand}

\begin{document}

\maketitle

\begin{abstract}
Phylogenetic networks provide a more general description of evolutionary relationships than rooted phylogenetic trees. One way to produce a phylogenetic network is to randomly place $k$ arcs between the edges of a rooted binary phylogenetic tree with $n$ leaves.  The resulting directed graph may fail to be a phylogenetic network, and even when it is 
it may fail to be a tree-child or normal network. In this paper, we first show that if $k$ is fixed, the proportion of arc placements  that result in a normal network tends to 1 as $n$ grows. From this result, the asymptotic enumeration of normal networks becomes straightforward and provides a transparent meaning to the combinatorial terms that arise.    Moreover, the approach  extends to allow $k$ to grow with $n$ (at the rate $o(n^\frac{1}{3})$), which was not handled in earlier work.  We also investigate a subclass of normal networks of particular relevance in biology (hybridization networks) and establish that the same asymptotic results apply.  
\end{abstract}



\section{Introduction}
Rooted phylogenetic networks (defined in Section~\ref{def} below) provide a precise way to represent the evolution of objects (species, viruses, languages etc.) under the twin processes of speciation and reticulation \cite{hus10}. The leaves (vertices of out-degree 0) of these networks typically correspond to observed individuals at the present, and the other vertices correspond to ancestral species.  Over the last two decades, the mathematical, statistical, and computational properties of phylogenetic networks have become an active area of research.  Various classes of networks with particular properties have been identified, and the relationships between various classes of networks has been investigated (for a recent survey, see \cite{kon22}).  
Exact and asymptotic enumeration techniques can then be used to determine the size of different network classes, and thereby compare the sizes of a class to a given subclass.    Combinatorial enumeration can also provides structural insight into various network classes, since such insights often arise as a byproduct of counting results.

The simplest phylogenetic network is a rooted tree, which models speciation (only). A slightly more general class is that of hybridization networks, which also allow pairs of species in the past to combine to form new (hybrid) species. 

In this paper, we focus on a class of networks that includes hybridization networks,  but is slightly more general, namely the class of {\em normal} networks. Such networks enjoy a number of desirable properties (see e.g., \cite{fra21}). We describe a new way to asymptotically count the class of normal networks with $n$ leaves and $k$ reticulation vertices  as $n$ becomes large (with $k$ fixed), a topic that has been investigated  by quite different methods by the first author in \cite{Fuchs et al. 2019}, \cite{Fuchs et al. 2021} and \cite{Fuchs et al. 2022}.  We then show how the results can be extended to allow $k$ to grow (slowly) with $n$, and then extend this approach to the subclass of hybridization networks. We begin with some definitions. 

\subsection{Definitions: Networks, decorated trees, and induced subdivision trees}\label{def}

In this paper, all trees and networks are directed graphs.
Throughout, we let $[n]$ denote the set $\{1, \dots, n\}$. A (binary) {\em phylogenetic network} on $[n]$ is a directed acyclic graph with $n$ leaves (vertices of out-degree 0) labelled bijectively by the elements of $[n]$, and with each non-leaf vertex having in-degree 1 and out-degree 2 (tree vertices) or in-degree 2 and out-degree 1 (reticulation vertices), or in-degree 0 and out-degree 1 (the root of the network at the top of an ancestral root edge). 
Edges which contain a reticulation vertex are called {\it reticulation edges}; all others are called {\it tree edges}. 

Two phylogenetic networks are regarded as equivalent if there is a directed graph isomorphism between them that maps $i$ to $i$ for each $i \in [n]$.  Three important  classes of phylogenetic networks are the following:

\begin{itemize}
    \item A {\em tree-child network}  is a phylogenetic network for which each non-leaf vertex has at least one of its outgoing edges directed to a tree vertex or a leaf.
    \item A {\em normal network} is a tree-child network that has no `shortcut' edge (i.e., no edge $(u,v)$ for which there is another path from $u$ to $v$).
        \item A {\em phylogenetic tree} is a phylogenetic network with no reticulation vertices.
\end{itemize}
Thus, tree-child networks include normal networks which include phylogenetic trees.
For more background and details on phylogenetic networks, see \cite{hus10}.

Let $\T_n$ denote the set of phylogenetic trees on leaf set $[n]$.
For $T\in \T_n$ and $k\geq 1$, let ${\mathcal S}(T,k)$ denote the set of all possible ordered pairs $(T_k, \omega_k)$, where $T_k$ and $\omega_k$ are defined recursively as follows: 
For $k=1$, set $\omega_1 = (p_1,p'_1)$, where $p_1$ subdivides some edge of $T$ and $p'_1$ subdivides some edge of the resulting tree. Let $T_1$ be the resulting subdivided tree (with two subdivision vertices). 

For $k>1$, let $\omega_k = \omega_{k-1} \cup \{(p_k, p'_k)\}$ where $p_k$ subdivides some edge of $T_{k-1}$ and $p'_k$ subdivides some edge of the resulting tree. Let $T_k$ be the resulting subdivided tree (with $2k$ subdivision vertices).

We call $(T_k, \omega_k)$ a {\em $k$-fold decorated tree on $[n]$} with base tree $T$.  Thus, $S(T, k)$ is the set of $k$-fold decorated trees with base tree $T$. 

Let $\G((T_k, \omega_k))$ be the directed graph obtained from $T_k$ by adding an arc from $p_i$ to $p_i'$ for each $i=1, \ldots, k$.  Note that $\G((T_k, \omega_k))$ may contain directed cycles (in particular, it need not be a phylogenetic network). 

\begin{figure}[htb]
\centering
\includegraphics[scale=1.0]{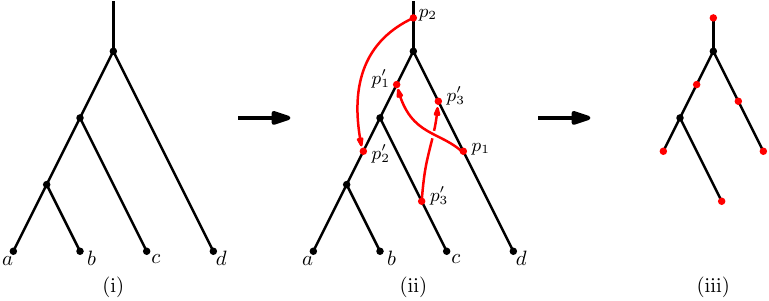}
\caption{(i): A phylogenetic tree $T$ with four leaves ($a,b,c,d \in [4]$), with all edges directed vertically downwards.  (ii) The directed graph $\G((T_3, \omega_3))$ obtained from $T$ by adding $k=3$ pairs of subdivision points to produce a $3$-fold decorated tree, and then joining $p_i$ to $p'_i$ for each $i$. In this example, the graph contains a directed cycle and thus does not correspond to a phylogenetic network. (iii) The associated induced subdivision tree which contains the six vertices $(p_1, p_2, p_3, p'_1, p'_2, p'_3$) (indicated in red)  and two other vertices of $T$ (indicated in black).}
\label{fig1}
\end{figure}

We also introduce a further notion associated with any $k$-fold decorated tree $(T_k,\omega_k)$. Consider the minimal subtree containing the $2k$ subdivision vertices, which we call the {\it induced subdivision tree}. Note that every leaf of the induced subdivision tree is a subdivision vertex  of $\G((T_k, \omega_k))$;  in addition, there might also be subdivision vertices which are non-leaf vertices, as occurs in Fig.~\ref{fig1}(iii).

These concepts are illustrated by the example shown in Fig.~\ref{fig1}. 

We also employ standard asymptotic notation:  $f(n) \sim g(n)$ if $\lim_{n \rightarrow \infty} \frac{f(n)}{g(n)}=1$, $f(n) = {\mathcal O}(g(n))$ if $f(n) \leq C g(n)$ for a constant $C$, and $f(n)=o(g(n))$ if $\lim_{n\rightarrow\infty}\frac{f(n)}{g(n)}=0$. Moreover, we use $[z^n]f(z)$ for the $n$-th coefficient of a generating function $f(z)$, and for any odd integer $n>1$, we let $n!!$ denote the product of the odd numbers from 1 to $n$ ($n$ `double factorial' or `semifactorial'). 

\bigskip

\section{Results}
We begin by counting the set ${\mathcal S}(T, k)$.
\begin{lemma}
\label{basic}
    The number of $k$-fold decorated trees $(T_k, \omega_k)$ on $[n]$ with base tree $T$ is given by:
    
    $$|{\mathcal S}(T, k)|=\frac{(2n-1)\cdot (2n)\cdots (2n+2k-2)}{k!}\sim \frac{4^kn^{2k}}{k!}.$$
\end{lemma}
\begin{proof}
The number of arcs in the tree $T$ (with an ancestral root edge) is $(2n-1)$ so there are this many choices for $p_1$. Placing $p_1$ creates a tree with $2n$ arcs, so there are $2n$ choices for $p'_1$.  Continuing in this way for the $k$ pairs of points $(p_i, p'_i)$ gives the term in the numerator, and the $k!$ in the denominator accounts for the fact that the same $k$-fold decorated tree can be obtained by placing the pairs of points $(p_i, p'_i)$ onto the arcs of $T$ in any order.

The asymptotic part of the result is obtained by noting that the numerator is a polynomial of degree $2k$ in $2n$.
\end{proof}

Next, we define the set:
\[
{\mathcal S}(n,k):=\bigcup_{T\in\T_n}\mathcal{S}(T,k).
\]
Thus, by Lemma~\ref{basic},
\[
\vert {\mathcal S}(n,k)\vert=\sum_{T\in\T_n}\vert {\mathcal S}(T,k)\vert=\frac{(2n-1)(2n)\cdots (2n+2k-2)}{k!}r_n.\,
\]
where
\[
r_n=|\T_n|=(2n-3)!!=\frac{(2n-2)!}{2^{n-1}(n-1)!}\sim\frac{\sqrt{2}}{2}\left(\frac{2}{e}\right)^n n^{n-1}.
\]
Consequently,
\begin{equation}\label{asymp-Snk}
\vert {\mathcal S}(n,k)\vert\sim\frac{2^{2k-1}\sqrt{2}}{k!}\left(\frac{2}{e}\right)^nn^{n+2k-1}.
\end{equation}
\begin{remarkX}\label{small-k}
For later purposes, we point out that (\ref{asymp-Snk}) also holds for $k=o(\sqrt{n})$ as
\begin{align*}
(2n-1)(2n)\cdots (2n+2k-2)&=(2n)^{2k}\left(1-\frac{1}{2n}\right)\prod_{j=0}^{2k-2}\left(1+\frac{j}{2n}\right)\\
&=(2n)^{2k}\left(1+{\mathcal O}(n^{-1})\right)e^{\sum_{j=0}^{2k-2}\log\left(1+\frac{j}{2n}\right)}\\
&=(2n)^{2k}\left(1+{\mathcal O}(n^{-1})\right)e^{\sum_{j=0}^{2k-2}{\mathcal O}(j/n)}\\
&=(2n)^{2k}\left(1+{\mathcal O}(n^{-1})\right)e^{{\mathcal O}(k^2/n)}\\
&=(2n)^{2k}\left(1+\mathcal{O}\left(\frac{k^2}{n}\right)\right),
\end{align*}
where we used:
\[
\sum_{j=0}^{2k-2}j=\frac{(2k-2)(2k-1)}{2}={\mathcal O}(k^2).
\]
\end{remarkX}

Next, we partition ${\mathcal S}(n,k)$ into three disjoint subsets:
\[
{\mathcal S}(n,k)={\mathcal S}_{c}(n,k) \sqcup {\mathcal S}_{no}(n,k) \sqcup {\mathcal S}_{\neg no}(n,k).
\]
The first set on the right (${\mathcal S}_{c}(n,k)$) consists of all $k$-fold decorated trees $(T_k,\omega_k)$ on $[n]$ such that $\G((T_k,\omega_k))$ contains a cycle. Thus the remaining $k$-fold decorated trees on $[n]$ are such that $\G((T_k,\omega_k))$ is a phylogenetic network. We then  partition this set of phylogenetic networks into two disjoint subsets: those networks for which $\G((T_k,\omega_k))$ is or is not a normal network (${\mathcal S}_{no}(n,k)$ and ${\mathcal S}_{\neg no}(n,k)$, respectively).

Our goal is to show that the contributions of ${\mathcal S}_c(n,k)$ and ${\mathcal S}_{\neg no}(n,k)$ are asymptotically negligible in (\ref{asymp-Snk}).
We start by observing the following.

\begin{lemma}\label{number-of-edges}
The induced subdivision tree has $4k-2-\ell$ edges where $\ell$ is the number of non-leaf subdivision vertices.
\end{lemma}
\begin{proof}
If the induced subdivision tree has $\ell$  non-leaf subdivision vertices, then it is a tree with $2k-\ell$ leaves, $\ell$ unary vertices (including possibly the root; see, for example, the third panel in Figure~\ref{fig1}), and the remaining $2k-\ell-1$ vertices are the binary vertices. Every vertex except the root has an incoming edge and this is the total number of edges. Thus, the number of edges is:
\[
\underbrace{2k-\ell}_{\text{leaves}}+\underbrace{\ell}_{\substack{\text{unary}\\ \text{vertices}}}+\underbrace{2k-\ell-1}_{\text{binary vertices}}-\underbrace{1}_{\text{root}}=4k-2-\ell.
\]
\end{proof}

In addition, we have the following.
\begin{lemma}\label{main-lemma}
Denote by $S_{n,k,\ell}$ the number of k-fold decorated trees $(T_k,\omega_k)$ on $[n]$ such that the induced subdivision tree has exactly $\ell$  non-leaf subdivision vertices. Then, for fixed $k$ and $\ell$, as $n\rightarrow\infty$,
\[
S_{n,k,\ell}={\mathcal O}\left(\left(\frac{2}{e}\right)^nn^{n+2k-1-\ell/2}\right).
\]
\end{lemma}
\begin{proof}
Consider induced subdivision trees with $2k-\ell$ leaves and $\ell$  non-leaf subdivision vertices. The $k$-fold decorated trees $(T_k, \omega_k)$ which are counted by $S_{n,k,\ell}$ are obtained from these induced subdivision trees by attaching phylogenetic trees below the leaves and replacing edges by  (ordered) sequences of phylogenetic trees suitably relabelled (which is automatically done by the use of exponential generating functions; see the sequence construction in Section II.2 of \cite{FlSe}), including an ancestral root edge into the root, which has to be added to the induced subdivision tree. Note that, by Lemma~\ref{number-of-edges}, there are $4k-1-\ell$ such sequences in total. 

The  construction above, in terms of  exponential generating function, gives the following for a fixed induced subdivision tree:
\begin{equation}\label{egf}
\underbrace{r(z)^{2k-\ell}}_{\substack{\text{trees below} \\ \text{leaves}}}\cdot\underbrace{\frac{1}{(1-r(z))^{4k-1-\ell}}}_{\substack{\text{sequences of trees}\\ \text{on edges}}},
\end{equation}
where
\[
r(z):=\sum_{n\geq 1}r_n\frac{z^n}{n!}=1-\sqrt{1-2z}
\]
counts phylogenetic trees and thus $1/(1-r(z))$ counts sequences of phylogenetic trees.

To obtain the coefficient of (\ref{egf}), we use singularity analysis; see Chapter VI of \cite{FlSe}. First, as $z\rightarrow 1/2$,
\[
\frac{r(z)^{2k-\ell}}{(1-r(z))^{4k-1-\ell}}\sim\frac{1}{(1-2z)^{2k-1/2-\ell/2}}.
\]
By Corollary~VI.1 in \cite{FlSe}, this gives (up to a constant) the upper bound for $S_{n,k,\ell}$
\[
n![z^n]\frac{r(z)^{2k-\ell}}{(1-r(z))^{4k-1-\ell}}\sim n!2^n\frac{n^{2k-3/2-\ell/2}}{\Gamma(2k-1/2-\ell/2)}={\mathcal O}\left(\left(\frac{2}{e}\right)^nn^{n+2k-1-\ell/2}\right)
\]
as claimed.
\end{proof}

We can now show that the contributions of ${\mathcal S}_c(n,k)$ are asymptotically negligible in (\ref{asymp-Snk}).

\begin{proposition}\label{cycles}
We have,
\[
\vert{{\mathcal S}_{c}(n,k)}\vert={\mathcal O}\left(\left(\frac{2}{e}\right)^n n^{n+2k-3/2}\right).
\]
\end{proposition}
\begin{proof}
If a $k$-fold decorated tree $(T_k,\omega_k)$ is in ${\mathcal S}_c(n,k)$, then $\G((T_k,\omega_k))$ contains a cycle. Thus, the induced subdivision tree has at least one non-leaf subdivision  vertex. Applying Lemma~\ref{main-lemma} gives the claim.
\end{proof}

Next, we show that the contribution of ${\mathcal S}_{\neg no}(n,k)$ is also negligible. 

\begin{proposition}\label{normal}
We have,
\[
\vert{\mathcal S}_{\neg no}(n,k)\vert={\mathcal O}\left(\left(\frac{2}{e}\right)^nn^{n+2k-3/2}\right).
\]
\end{proposition}
\begin{proof}
By Lemma~\ref{main-lemma}, we can restrict ourselves to $k$-fold decorated trees $(T_k,\omega_k)$ whose induced subdivision tree has $2k$ leaves. Note that in these decorated trees, the network $\G((T_k,\omega_k))$ has trees below the reticulation vertices and thus does not contain a reticulation vertex followed by a reticulation vertex. In addition, if $\G((T_k,\omega_k))$ contains a tree vertex followed by two reticulation vertices, then the number of such decorated trees is (up to a constant) bounded by
\[
n![z^n]\frac{r(z)^{2k}}{(1-r(z))^{4k-1-2}};
\]
see the explanation in the proof of Lemma~\ref{main-lemma} (where we now have $\ell=0$). Here, the additional $-2$ in the exponent of the denominator arises from the two edges below the above tree vertex in $\G((T_k,\omega_k))$ as empty sequences of phylogenetic trees are attached to these edges in the induced subdivision tree of $(T_k,\omega_k)$. This bound is
\begin{equation}\label{two-edges}
n![z^n]\frac{r(z)^{2k}}{(1-r(z))^{4k-1-2}}={\mathcal O}\left(\left(\frac{2}{e}\right)^nn^{n+2k-2}\right).
\end{equation}
Likewise, if we have a shortcut in $\G((T_k,\omega_k))$ (which violates the normal condition), then we obtain the upper bound
\begin{equation}\label{one-edge}
n![z^n]\frac{r(z)^{2k}}{(1-r(z))^{4k-1-1}}={\mathcal O}\left(\left(\frac{2}{e}\right)^nn^{n+2k-3/2}\right),
\end{equation}
where the additional $-1$ comes from the empty sequence attached to the shortcut in the induced subdivision tree of $(T_k,\omega_k)$ (which must be part of the induced subdivision tree). Combining these two upper bounds gives the claimed result.
\end{proof}

Propositions~\ref{cycles} and \ref{normal} provide an alternative and immediate way to asymptotically count normal networks with a given number of reticulations ($k$). Although this result is known (from \cite{Fuchs et al. 2019}, \cite{Fuchs et al. 2021}, \cite{Fuchs et al. 2022}) our proof here provides a more transparent way to see why the asymptotic result holds; more importantly, it can be extended to allow $k$ to grow (slowly) with $n$ (as we describe in the next section), unlike the earlier approaches. 

Let $N_{n,k}$ denote the number of normal  phylogenetic networks with $n$ leaves and $k$ reticulation vertices.
    
  
\begin{corollary}
For fixed $k$, as $n\rightarrow\infty$,
\begin{equation}
\label{eq1}
    N_{n,k} \sim \frac{2^{k-1}\sqrt{2}}{k!}\left(\frac{2}{e}\right)^n n^{n+2k-1}.
\end{equation}
\end{corollary}

\begin{proof}
Let $L_n$ be the number of pairs $(N, T)$ where $N$ is a normal network with leaf set $[n]$ and $k$ reticulation vertices, and $T \in\T_n$ is displayed by $N$  (i.e. $T$ can be  obtained by the following process: For each reticulation vertex $v$ of $N$ delete one of the two incoming arcs and suppress the resulting subdivision vertex).  Since any normal network with $k$ reticulation vertices displays exactly $2^k$ distinct phylogenetic trees (Corollary 3.4 of \cite{Willson 2012}), we have:
\begin{equation}
\label{eq3}
L_n = 2^k \cdot N_{n,k}.
\end{equation}
In addition, by the definition of ${\mathcal S}(n,k)$, we have
\[
L_n=\vert{\mathcal S}_n(n,k)\vert.
\]
Now, from (\ref{asymp-Snk}) and Propositions~\ref{cycles} and ~\ref{normal}, we have:
\[
L_n\sim \frac{2^{2k-1}\sqrt{2}}{k!}\left(\frac{2}{e}\right)^n n^{n+2k-1}
\]
and so, by (\ref{eq3}):
\[
N_{n,k}\sim\frac{2^{k-1}\sqrt{2}}{k!}\left(\frac{2}{e}\right)^nn^{n+2k-1},
\]
which establishes~(\ref{eq1}), as required.
\end{proof}

\bigskip

\section{Allowing $k$ to grow (slowly) with $n$}\label{dep-on-k}

In order to understand the range of validity of (\ref{eq1}) when $k$ is allowed to grow with $n$, we have to make the dependence on $k$ in the $\mathcal{O}$ term in Proposition~\ref{cycles} and Proposition~\ref{normal} explicit. Since both of these propositions crucially depend on Lemma~\ref{main-lemma}, we first revisit the proof of this lemma.

From now on, we assume that $k=o(n^{1/3})$. This will turn out to be the range of $k$ for which we can show that (\ref{eq1}) is still valid.

By the proof of Lemma~\ref{main-lemma}, the number of $k$-fold decorated trees $(T_k,\omega_k)$ on $[n]$ with a {\it fixed} induced subdivision tree having $2k-\ell$ leaves and $\ell$ non-leaf subdivision vertices is bounded by
\begin{equation}\label{up-dec-trees}
n![z^n]\frac{r(z)^{2k-\ell}}{(1-r(z))^{4k-1-\ell}}={\mathcal O}\left(\left(\frac{2}{e}\right)^nn^{n+2k-1-\ell/2}\right).
\end{equation}
Our first goal is to sharpen this to
\begin{equation}\label{ub-main-lemma}
n![z^n]\frac{r(z)^{2k-\ell}}{(1-r(z))^{4k-1-\ell}}={\mathcal O}\left(\left(\frac{2}{e}\right)^n\frac{n^{n+2k-1-\ell/2}}{\Gamma(2k-1/2-\ell/2)}\right),
\end{equation}
where the implied constant in ${\mathcal O}$ is absolute (i.e., it does not dependent on $\ell,k,$ and $n$). In this section, ${\mathcal O}$ will always be assumed to have this property.

In order to prove this, we start with a lemma.
\begin{lemma}
For $\alpha>0$ with $\alpha=o(\sqrt{n})$, we have the following uniform bound
\[
[z^n](1-z)^{-\alpha}={\mathcal O}\left(\frac{n^{\alpha-1}}{\Gamma(\alpha)}\right).
\]
\end{lemma}
\begin{proof}
By the binomial theorem,
\begin{equation}\label{binom-thm}
[z^n](1-z)^{-\alpha}=\binom{n+\alpha-1}{n}=\frac{\Gamma(n+\alpha)}{\Gamma(n+1)\Gamma(\alpha)}.
\end{equation}
Next, by Stirling's formula for the Gamma function
\[
\Gamma(x)=\left(\frac{x}{e}\right)^x\sqrt{\frac{2\pi}{x}}\left(1+{\mathcal O}\left(\frac{1}{x}\right)\right),\qquad (x\rightarrow\infty),
\]
we have:
\begin{align*}
\log\frac{\Gamma(n+\alpha)}{\Gamma(n+1)}&=\log\Gamma(n+\alpha)-\log\Gamma(n+1)\\
&=(n+\alpha)\log(n+\alpha)-(n+\alpha)-\frac{1}{2}\log(n+\alpha)+\frac{1}{2}\log(2\pi)\\
&\quad-(n+1)\log(n+1)+n+1+\frac{1}{2}\log(n+1)-\frac{1}{2}\log(2\pi)+{\mathcal O}(n^{-1})\\
&=(\alpha-1)\log n+(n+\alpha-1/2)\log(1+\alpha/n)-\alpha+{\mathcal O}(n^{-1})\\
&=(\alpha-1)\log n+{\mathcal O}\left(\frac{\alpha^2+1}{n}\right).
\end{align*}
Thus,
\[
\frac{\Gamma(n+\alpha)}{\Gamma(n+1)}=n^{\alpha-1}\left(1+{\mathcal O}\left(\frac{\alpha^2+1}{n}\right)\right)
\]
and combining this with (\ref{binom-thm}) yields the claim.
\end{proof}

Now, to establish (\ref{ub-main-lemma}), we use $r(z)=1-\sqrt{1-2z}$ and the binomial theorem:
\[
[z^n]\frac{r(z)^{2k-\ell}}{(1-r(z))^{4k-1-\ell}}=\sum_{j=0}^{2k-\ell}\binom{2k-\ell}{j}(-1)^j[z^n](1-2z)^{-2k+1/2+\ell/2+j/2}.
\]
By the lemma, the $j$-th term becomes
\begin{align*}
{\mathcal O}&\left(\binom{2k-\ell}{j}2^{n}\frac{n^{2k-3/2-\ell/2-j/2}}{\Gamma(2k-1/2-\ell/2-j/2)}\right)\\
&\qquad={\mathcal O}\left(\frac{2^nn^{2k-3/2-\ell/2}}{\Gamma(2k-1/2-\ell/2)}\frac{\Gamma(2k-1/2-\ell/2)}{\Gamma(2k-1/2-\ell/2-j/2)}\frac{(2k/\sqrt{n})^j}{j!}\right)\\
&\qquad={\mathcal O}\left(\frac{2^n n^{2k-3/2-\ell/2}}{\Gamma(2k-1/2-\ell/2)}\frac{(2k/\sqrt[3]{n})^{3j/2}}{j!}\right),
\end{align*}
where, in the second-last step, we used:
\[
\frac{\Gamma(2k-1/2-\ell/2)}{\Gamma(2k-1/2-\ell/2-j/2)}={\mathcal O}((2k)^{j/2}),
\]
which follows from standard estimates for the ratio of gamma functions. Summing the above estimate over $j$ gives
\begin{align*}
[z^n]\frac{r(z)^{2k-\ell}}{(1-r(z))^{4k-1-\ell}}&={\mathcal O}\left(\frac{2^nn^{2k-3/2-\ell/2}}{\Gamma(2k-1/2-\ell/2)}e^{(2k/\sqrt[3]{n})^{3/2}}\right)\\
&={\mathcal O}\left(\frac{2^nn^{2k-3/2-\ell/2}}{\Gamma(2k-1/2-\ell/2)}\right).
\end{align*}
Multiplying this by the asymptotics of $n!$ gives (\ref{ub-main-lemma}).

Next, we have to multiply (\ref{ub-main-lemma}) by the number of induced subdivision trees with $2k-\ell$ leaves and $\ell$ non-leaf subdivision vertices which is given by: 
\[
\binom{2k}{\ell} r_{2k-\ell}\frac{(4k-2\ell-1)\cdots (4k-\ell-2)}{k!}
\]
as they are enumerated by starting with a phylogenetic tree on $2k-\ell$ leaves, choosing $\ell$ non-leaf subdivision vertices on the edges of this tree, and redistributing the labels. Note that
\[
\binom{2k}{\ell}r_{2k-\ell}\frac{(4k-2\ell-1)\cdots(4k-\ell-2)}{\Gamma(2k-1/2-\ell/2)k!}\leq\frac{4^{\ell}k^{\ell}}{\ell!k!}\frac{\Gamma(2k+1)}{\Gamma(2k-1/2-\ell/2)}\frac{C_{2k-\ell-1}}{2^{2k-\ell-1}}.
\]
Here, we used that
\[
(4k-2\ell-1)\cdots(4k-\ell-2)\leq 4^{\ell}k^{\ell}
\]
and $r_n=n!C_{n-1}/2^{n-1}$, where $C_n$ denotes the $n$-th Catalan number:
\[
C_n=\frac{1}{n+1}\binom{2n}{n}\sim\frac{4^n}{\sqrt{\pi n^3}}.
\]
Therefore, the above bound becomes:
\begin{equation}\label{coef-est}
\frac{4^{\ell}k^{\ell}}{\ell!k!}\frac{\Gamma(2k+1)}{\Gamma(2k-1/2-\ell/2)}\frac{C_{2k-\ell-1}}{2^{2k-\ell-1}}={\mathcal O}\left(\frac{4^k(2k)^{3\ell/2}k^{3/2}}{\ell!k!(2k-\ell)^{3/2}}\right),
\end{equation}
where we used
\[
\frac{\Gamma(2k+1)}{\Gamma(2k-1/2-\ell/2)}={\mathcal O}\left(2^{\ell/2}k^{(\ell+3)/2}\right).
\]
Combining (\ref{ub-main-lemma}) and (\ref{coef-est}), the bound in Lemma~\ref{main-lemma} can be sharpened to
\[
S_{n,k,\ell}={\mathcal O}\left(\frac{(2k)^{3\ell/2}k^{3/2}}{\ell!(2k-\ell)^{3/2}}\frac{4^k}{k!}\left(\frac{2}{e}\right)^n n^{n+2k-1-\ell/2}\right).
\]

The bound in Proposition~\ref{cycles} for $\vert {\mathcal S}_{c}(n,k)\vert$ is obtained by summing the bound for $S_{n,k,\ell}$ for $\ell$ from $1$ to $2k-1$. Consequently,
\[
\vert{\mathcal S}_{c}(n,k)\vert={\mathcal O}\left(c(n,k)\frac{4^k}{k!}\left(\frac{2}{e}\right)^n n^{n+2k-1}\right),
\]
where
\[
c(n,k):=k^{3/2}\sum_{\ell=1}^{2k-1}\frac{(2k/\sqrt[3]{n})^{3\ell/2}}{\ell!(2k-\ell)^{3/2}}
\]
We break the above sum into two parts according to whether $\ell<k$ or $\ell\geq k$. For the first part, we get
\[
k^{3/2}\sum_{\ell=1}^{k-1}\frac{(2k/\sqrt[3]{n})^{3\ell/2}}{\ell!(2k-\ell)^{3/2}}={\mathcal O}\left(\sum_{\ell=1}^{k-1}(2k/\sqrt[3]{n})^{3\ell/2}\right)={\mathcal O}\left((2k/\sqrt[3]{n})^{3/2}\right)=o(1),
\]
where, in the last step, we used that $k=o(n^{1/3})$. For the second part, we have
\[
k^{3/2}\sum_{\ell=k}^{2k-1}\frac{(2k/\sqrt[3]{n})^{3\ell/2}}{\ell!(2k-\ell)^{3/2}}={\mathcal O}\left(k^{3/2}(2k/\sqrt[3]{n})^{3k/2}\right)=o(1),
\]
where the last step is clear for bounded $k$ and holds for $k\rightarrow\infty$ because $4k\leq\sqrt[3]{n}$ for large $k$ (again by $k=o(n^{1/3})$). Thus, we have
\begin{equation}\label{est-Sc}
\vert{\mathcal S}_c(n,k)\vert=o\left(\frac{4^k}{k!}\left(\frac{2}{e}\right)^n n^{n+2k-1}\right)
\end{equation}
for our range of $k$.

Next, we consider the estimate of Proposition~\ref{normal} which, as explained in the proof, was obtained by estimating separately the number of $k$-fold decorated trees $(T_k,\omega_k)$ on $[n]$ with $\G((T_k,\omega_k))$ a phylogenetic network whose induced subdivision tree has (i) at least one non-leaf subdivision vertex or otherwise (ii) two outgoing edges of a tree node where no sequence is attached to or (iii) one edge where no sequence is attached to in the induced subdivision tree. 

The first case is treated as above. For the second and third cases, we use the estimate (\ref{two-edges}) and (\ref{one-edge}) instead of (\ref{up-dec-trees}), where both bounds have to be multiplied by $k$ which is the upper bound of the number of tree nodes and edges in the induced subdivision tree, respectively. Moreover, we can use the same constant as in (\ref{coef-est}) but with $\ell=0$ and multiplied by $\Gamma(2k-1/2)/\Gamma(2k-3/2)={\mathcal O}(k)$ in the second case and by $\Gamma(2k-1/2)/\Gamma(2k-1)={\mathcal O}(k^{1/2})$ in the third case. Overall, for the second case, this gives
\[
{\mathcal O}\left(\frac{k^2}{n}\frac{4^k}{k!}\left(\frac{2}{e}\right)^n n^{n+2k-1}\right)=o\left(\frac{4^k}{k!}\left(\frac{2}{e}\right)^n n^{n+2k-1}\right)
\]
and for the third case
\begin{equation}\label{case3}
{\mathcal O}\left(\frac{k^{3/2}}{\sqrt{n}}\frac{4^k}{k!}\left(\frac{2}{e}\right)^n n^{n+2k-1}\right)=o\left(\frac{4^k}{k!}\left(\frac{2}{e}\right)^n n^{n+2k-1}\right)
\end{equation}
Combining these bounds with the bound for the first case, we can improve the result of Proposition~\ref{normal} to
\begin{equation}\label{est-Snn}
\vert{\mathcal S}_{\neg no}(n,k)\vert=o\left(\frac{4^k}{k!}\left(\frac{2}{e}\right)^n n^{n+2k-1}\right)
\end{equation}
for $k=o(n^{1/3})$.

Now, by combining (\ref{est-Sc}), (\ref{est-Snn}), and (\ref{asymp-Snk}) (which holds for our range of $k$; see Remark~\ref{small-k}), we see that (\ref{eq1}) holds even if $k$ is allowed to grow moderately with $n$, namely, for $k=o(n^{1/3})$.

\section{Hybridization networks}
A {\em hybridization network}  is a tree-child network on leaf set $X$, which has at least one temporal ordering (or `ranking'). This means that one can assign a real-valued temporal date ($T(v)$) to each vertex $v$ of the network so that (i) if $(u,v)$ is a tree edge then $T(u)<T(v)$ and (ii) if $v$ is a reticulation vertex with parents $u$ and $w$ then $T(u)=T(v)=T(w)$. Hybridization networks are particularly relevant to biology, since they model species' evolution that comprises two processes: binary speciation events (as in phylogenetic trees), and events where two contemporaneous species hybridize to give rise to a new (hybrid) species (see e.g., \cite{mar14}).  

\begin{figure}[htb]
\centering
\includegraphics[scale=0.6]{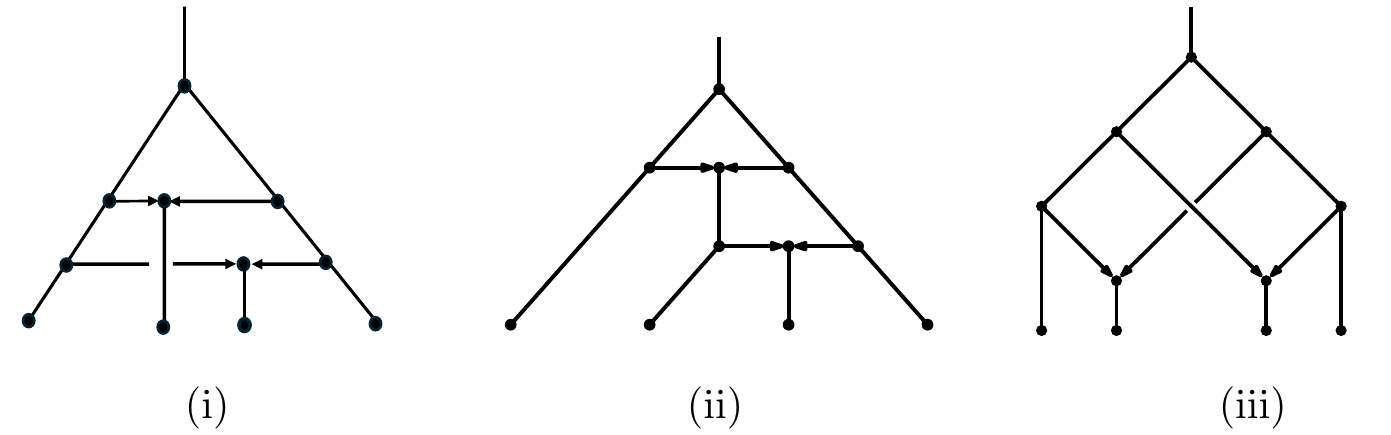}
\caption{The  three shapes of normal networks on $n=4$ leaves with $k=2$ reticulation vertices. Cases (i) and (ii) correspond to hybridization networks, and Case (iii) corresponds to a normal network that is not a hybridization network. There are  12 distinct phylogenetic networks of shape (i), 24 of shape (ii), and 12  of shape (iii). }
\label{fig2}
\end{figure}

It can easily be shown that every hybridization network is normal (see e.g., Proposition 10.12 of \cite{ste16}); however, the converse does not hold, as Fig.~\ref{fig2} shows. 

Hybridization networks also correspond to the class of tree-child networks that can be `ranked' in the sense described in \cite{bie22}. However, although there is a simple, exact, and explicit formula for counting ranked tree-child networks with $n$ leaves and $k$ reticulation vertices, the exact enumeration of hybridization networks is more complex. 

Let $H_{n,k}$ denote the number of hybridization networks with $n$ leaves and $k$ reticulation vertices. $H_{n,1}$ is just the number of normal networks with one reticulation vertex, and so $H_{4,1}=54$; see (\ref{Hn1}). In addition,  $H_{4,2}=36$, compared with the 48 possible normal networks with $n=4, k=2$ (see Fig.~\ref{fig2}).

\subsection{Enumeration of $H_{n,k}$ for $k=1,2$}
\label{hnksec}

$H_{n,1}$ coincides with the number of normal networks with one reticulation vertex and thus:
\begin{equation}\label{Hn1}
H_{n,1}=N_{n,1}= \frac{1}{2}n![z^n]\left(\frac{r(z)}{1-r(z)}\right)^3 = \frac{1}{2}[(2n+1)!!+3(2n-1)!!]-3n!2^{n-1},
\end{equation}
where $r(z) = 1-\sqrt{1-2z}$; see \cite{Zh} or \cite{Fuchs et al. 2021}.

For $k=2$, we have the following result.

\begin{proposition}
\label{proH} We have,
    $$H_{n,2} = (2n-1)!!(n^3+9n^2-16n-12)-3n!2^n(n^2-4).$$
\end{proposition}

\begin{proof}
The proof uses the decomposition of a network into {\it bridgeless components} from \cite{BoGaMa}. First, we recall the definition of a bridgeless component from graph theory: a bridgeless component of a graph $G$ is a maximal induced subgraph of $G$ without cut edges (bridges). Let $\mathcal{H}_2$ denote the set of hybridization networks with exactly 2 reticulation vertices. 
Consider the exponential generating function for $H_{n,k}$ defined by:
$$H_k(z) = \sum_{n=1}^{\infty} H_{n, k}\frac{z^n}{n!}.$$ Any hybridization network $N \in \mathcal{H}_2$ satisfies exactly one of the following cases (for details, we refer to the Appendix):

Case 1: The root of $N$ belongs to a bridgeless component which contains 0 reticulation vertices. This contributes $H_2(z)r(z)+
    \frac{H_1(z)^2}{2}$ to the exponential generating function of $\mathcal{H}_2$.

Case 2: The root of $N$ belongs to a bridgeless component which contains exactly one reticulation vertex. This contributes $H_1(z)\frac{r(z)^2}{(1-r(z))^3}+\frac{1}{2}H_1(z)\frac{r(z)^2}{(1-r(z))^2}$ to the exponential generating function of $\mathcal{H}_2$.

 Case 3: The root of $N$ belongs to a bridgeless component which contains exactly two reticulation vertices. This contributes $\frac{1}{2}\frac{r(z)^4}{(1-r(z))^4}+\frac{r(z)^4}{(1-r(z))^5}+\frac{7}{2}\frac{r(z)^5}{(1-r(z))^5}+\frac{5}{4}\frac{r(z)^6}{(1-r(z))^6}$ to the exponential generating function of $\mathcal{H}_2$.
 
Then, for $k=2$, we have:
\begin{align*}
H_2(z)&=H_2(z)r(z)+\frac{H_1(z)^2}{2}+H_1(z)\frac{r(z)^2}{(1-r(z))^3}+\frac{1}{2}H_1(z) \frac{r(z)^2}{(1-r(z))^2}\\
&\quad+\frac{1}{2}\frac{r(z)^4}{(1-r(z))^4}+\frac{r(z)^4}{(1-r(z))^5}+ \frac{7}{2}\frac{r(z)^5}{(1-r(z))^5}+\frac{5}{4}\frac{r(z)^6}{(1-r(z))^6} 
\end{align*}
and consequently,
\begin{align*}
H_2(z)&=\frac{H_1(z)^2}{2(1-r(z))}+H_1(z)\frac{r(z)^2}{(1-r(z))^4}+\frac{1}{2}H_1(z) \frac{r(z)^2}{(1-r(z))^3}\\
&\quad+\frac{1}{2}\frac{r(z)^4}{(1-r(z))^5}+\frac{r(z)^4}{(1-r(z))^6}+ \frac{7}{2}\frac{r(z)^5}{(1-r(z))^6}+\frac{5}{4}\frac{r(z)^6}{(1-r(z))^7}, 
\end{align*}
where (from (\ref{Hn1})):
\[
H_1(z)=\frac{1}{2}\frac{r(z)^3}{(1-r(z))^3}.
\]

$H_2(z)$ can be rewritten as:
\begin{align*}
    H_2(z)&=\frac{15}{8}(1-r(z))^{-7}-6(1-r(z))^{-6}+\frac{27}{8}(1-r(z))^{-5}+9(1-r(z))^{-4}\\&\quad-\frac{123}{8}(1-r(z))^{-3}+9(1-r(z))^{-2}-\frac{15}{8}(1-r(z))^{-1}.
\end{align*}

By using this equation, it can be shown that:
$$H_{n,2}=n![z^n]H_2(z)= (2n-1)!!(n^3+9n^2-16n-12)-3n!2^{n}(n^2-4).$$
This proves the claim.
\end{proof}

By Proposition~\ref{proH}, we have $H_{4,2}=36, H_{5,2} = 1890$, and $H_{6,2}=66960$.

By comparison, the number $N_{n,2}$ of  normal networks with two reticulation vertices is given (from \cite{Fuchs et al. 2021}) by:
$$N_{n,2} = \frac{1}{3}(2n-1)!!(3n-4)(n^2+11n+6) -n!2^n(3n^2+2n-8).$$

It is readily verified that $H_{n,2}/N_{n,2} \sim 1$ as $n \rightarrow \infty.$
Thus it is of interest to consider the (asymptotic) number of normal networks with two reticulation vertices that are not hybridization networks.

From the above expressions we have:
$$N_{n,2}-H_{n,2} \sim \frac{2\sqrt{2}}{3}\left(\frac{2}{e}\right)^nn^{n+2}, \quad \text{as} \quad n \to \infty.$$

\subsection{Asymptotics of $H_{n,k}$} 

We start with a useful notion. We call two reticulation edges of a phylogenetic network {\it collinear} if a vertex of one of them is connected by a tree path (i.e. a path consisting just of tree edges) to a vertex of the other one. For example, in Fig.~\ref{fig2}(iii), the two parallel reticulation edges that slope downwards to the right are collinear. 

Using this notion, we have the following result.

\begin{lemma}
\label{lemcol}
Let $(T_k,\omega_k)$ be a $k$-fold decorated tree on $[n]$ such that $\G((T_k,\omega_k))$ is a normal network which has no collinear reticulation edges. Then, $\G((T_k,\omega_k))$ is a hybridization network.
\end{lemma}
\begin{proof}
The proof proceeds by induction on $k$. The claim obviously holds for $k=0$, as, in this case, $\G((T_k,\omega_k))$ is just a tree, and any tree is a hybridization network. 

Now suppose that the claim holds for $k-1$; we are going to establish it for $k$. Pick a reticulation vertex of $\G((T_k,\omega_k))$. Note that, by assumption, the three subgraphs induced by the descendant set of the reticulation vertex as well as the descendant sets of the two parents of the reticulation vertex are all trees. Remove them together with the reticulation vertex from $\G((T_k,\omega_k))$. The remaining structure corresponds to a $(k-1)$-fold decorated tree $(T_{k-1}',\omega_{k-1}')$ on a set of leaves that consist of a subset of $[n]$ together with two new leaves that correspond to the parents of the removed reticulation vertex and which receive new labels (say, $n+1, n+2$). By applying the induction hypothesis, we obtain that $\G((T_{k-1}',\omega_{k-1}'))$ is a hybridization network, i.e., it can be drawn from top to bottom such that its events are in chronological order. Next, find the two extant lineages leading to $n+1$ and $n+2$ and attach the deleted reticulation event, which becomes the next event in the temporal order. Moreover, attach to the leaves with label $n+1$ and $n+2$ the two deleted subtrees and below the re-attached reticulation vertex the third deleted subtree. Clearly, the events of the three subtrees can be ordered such that $\G((T_k,\omega_k))$ becomes a hybridization network. This proves the claim. 
\end{proof}

Now, consider the set ${\mathcal S}_{no}(n,k)$, which we partition into the set which contains hybridization networks and the set which does not:
\[
{\mathcal S}_{no}(n,k)={\mathcal S}_{hyb}(n,k)\sqcup {\mathcal S}_{\neg hyb}(n,k).
\]
The cardinality of the set of hybridization networks again satisfies the asymptotics in (\ref{asymp-Snk}) since we have the following result.

\begin{proposition}\label{hybrid}
We have,
\[
\vert{\mathcal S}_{\neg hyb}(n,k)\vert={\mathcal O}\left(\left(\frac{2}{e}\right)^n n^{n+2k-3/2}\right).
\]
\end{proposition}
\begin{proof}
As in the proof of Proposition~\ref{normal}, we can restrict ourselves to $k$-fold decorated trees $(T_k,\omega_k)$ whose induced subdivision tree has $2k$ leaves. In addition, if $(T_k,\omega_k)\in{\mathcal S}_{\neg hyb}(n,k)$, then, by Lemma~\ref{lemcol}, $\G((T_k,\omega_k))$ has collinear reticulation edges. Clearly, if one starts from an induced subdivision tree with $2k$ leaves, such a network can only be obtained from it (using the procedure from the proof of Lemma~\ref{cycles}) by leaving at least one edge empty (i.e., by not attaching a sequence of subtrees to it). Thus, this situation is akin to the case of shortcuts in the proof of Proposition~\ref{normal}, which gave the bound (\ref{one-edge}). Since this is also the bound claimed in the current result, we are finished.
\end{proof}

In summary, we have the following result.
\begin{theorem}
For fixed $k$, as $n\rightarrow\infty$,
\begin{equation}\label{asymp-Hnk}
H_{n,k}\sim\frac{2^{k-1}\sqrt{2}}{k!}\left(\frac{2}{e}\right)^n n^{n+2k-1}.
\end{equation}
In addition, this asymptotic result is still valid in the range $k=o(n^{1/3})$.
\end{theorem}
\begin{proof}
The claimed expansion (\ref{asymp-Hnk}) follows from (\ref{asymp-Snk}) combined with Proposition~\ref{cycles}, Proposition~\ref{normal}, and Proposition~\ref{hybrid}. Moreover, that (\ref{asymp-Hnk}) holds for $k=o(n^{1/3})$ is proved by making the dependence on $k$ in the constant of Proposition~\ref{hybrid} explicit, which is handled as in Section~\ref{dep-on-k} (as explained in the proof of Proposition~\ref{hybrid}, the situation is akin to the case of shortcuts in the proof of Proposition~\ref{normal}, which gave the bound (\ref{case3})).    
\end{proof}

\section{Concluding comments}
In this paper, we have investigated the enumerative aspects of  constructing a phylogenetic network by placing arcs between the edges of the tree. Provided that the number of arcs placed ($k$) is constant or grows slowly enough with the number of leaves ($n$), the directed graph we construct is almost surely a 
phylogenetic network and, in addition, it (almost surely) belongs to the much smaller class of normal networks.  This result, combined with a combinatorial property of normal networks, allows an asymptotic enumeration of this class. Our approach provides an explicit interpretation of the various terms in the asymptotic formula, and  extends earlier results by allowing $k$ to depend on $n$. We also show that the same asymptotic results apply for the even smaller subclass of hybridization networks.  

Our analysis requires $k$ to grow no faster than $o(n^{1/3})$ and a natural question is whether the bound $k= o(n^{1/3})$ might be improved. For example, $k=o(n^{1/2})$ seems to be the range where the corresponding result for tree-child networks holds; see \cite{PoBa}. Is the same true for normal and/or hybridization networks? Or do they behave differently?

We end with some general observations. First, it is well known that any normal network has at most $n-2$ reticulation vertices, and  by Theorem 5.1 of \cite{mcd15}, almost all normal networks with $n$ leaves have $(1+o(1))n$ reticulation vertices. Thus the two classes of networks we are enumerating are not representative of normal networks selected uniformly at random. Nor are our normal networks representative of a randomly chosen tree-child network with $n$ leaves, since the proportion of the latter that are normal tends to 0 as $n$ grows (again by results in \cite{mcd15}). Nevertheless, almost all of the tree-child networks (with $k$ (fixed) reticulation vertices) that arise under the process we study will be normal networks (since this class of tree-child networks follows the same asymptotic law \cite{Fuchs et al. 2019}).  

\section{Acknowledgments} This research was carried out during a sabbatical stay of MF at the Biomathematics Research Center, University of Canterbury, Christchurch. He thanks the department and MS for hospitality, and the NSTC, Taiwan (NSTC-113-2918-I-004-001 and NSTC-113-2115-M-004-004-MY3) for financial support. MS and QZ thank the NZ Marsden Fund for research support (23-UOC-003). We also thank the two anonymous reviewers for their helpful  comments and suggestions.

\section{Appendix}
It was mentioned in Section~\ref{hnksec} that any hybridization network $N \in\mathcal{H}_{2}$ satisfies exactly one of the following three cases: (1) The root of $N$ belongs to a bridgeless component which contains 0 reticulation vertices; (2) The root of $N$ belongs to a bridgeless component which contains exactly 1 reticulation vertex; (3) The root of $N$ belongs to a bridgeless component which contains exactly 2 reticulation vertices. Each case makes corresponding contributions to the exponential generating function of $\mathcal{H}_2$.

In this section, we will explain more carefully how to derive the exponential generating function of $\mathcal{H}_2$ by considering these three cases; we will follow the approach from \cite{BoGaMa}. 

A hybridization network $N$ is said to be $level\text{-}k$ if the number of reticulation vertices contained in any bridgeless component of $N$ is at most $k$. 

For any bridgeless component $B$ with $k_B \leq k$ reticulation vertices of a level-$k$ hybridization network $N$, there exist at least two bridges of $N$ attached to $B$ because without this restriction, such networks would have an unbounded number of internal vertices resulting in an infinite number of these networks.

For an arbitrary hybridization network $N \in \mathcal{H}_2$, considering the bridgeless component containing the root $\rho$ denoted as $B_{\rho}$, suppressing any vertices of in-degree and out-degree 1, the resulting directed multi-graph is called a {\it generator}. A generator induced from $B_{\rho}$ may have 0, 1, or 2 reticulation vertex (vertices) and is called a {\it level-}$0$, {\it level-}$1$ or {\it level-}$2$ generator. We call non-terminal edges and vertices of out-degree 0 of a level-$k$ generator {\it sides}. For each level, there is a finite number of generator(s). Therefore, the three cases can be considered as: (1) The root of $N$ belongs to a level-$0$ generator; (2) The root of $N$ belongs to a level-$1$ generator; (3) The root of $N$ belongs to a level-$2$ generator.

We will now consider these three cases. Let $\mathcal{H}_1$ denote the set of hybridization networks with exactly 1 reticulation vertex, and $\mathcal{H}_0$ denote the set of hybridization networks with 0 reticulation vertices.

\subsection{Case 1}
\begin{figure}[htb]
\centering
\includegraphics[scale=1.0]{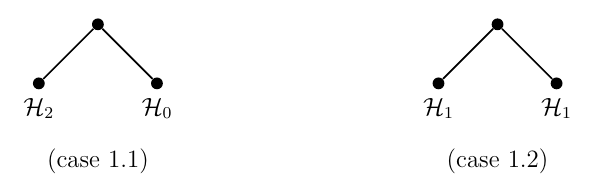}
\caption{(case 1.1): A binary root vertex with two children that are roots of networks of $\mathcal{H}_2$ and $\mathcal{H}_0$. (case 1.2): A binary root vertex with two children that are roots of networks from $\mathcal{H}_1$ whose left-to-right order is irrelevant.}
\label{fig_case1}
\end{figure}

For any $N \in \mathcal{H}_2$ such that the root of $N$ belongs to a level-$0$ generator, the root has two children. They satisfy one of two subcases:

\begin{itemize}
    \item One child is the root of a network from $\mathcal{H}_2$ and the other child is the root of a network from $\mathcal{H}_0$, see case 1.1 in Fig.~\ref{fig_case1}. This contributes $H_2(z)r(z)$ to the exponential generating function of $\mathcal{H}_2$.
    \item Both children are the root of networks of $\mathcal{H}_1$; see case 1.2 in Fig.~\ref{fig_case1}. This contributes $\frac{H_2(z)^2}{2}$ to the exponential generating function of $\mathcal{H}_2$. Here, the factor $1/2$ is because the left-to-right order is irrelevant.
\end{itemize}

Overall, Case 1 contributes $H_2(z)r(z)+\frac{H_2(z)^2}{2}$ to the exponential generating function of $\mathcal{H}_2$.

\subsection{Case 2}
Any $N \in \mathcal{H}_2$ such that the root of $N$ belongs to a level-$1$ generator satisfies one of two subcases:

\begin{itemize}
\item The child of the reticulation vertex is the root of a network from $\mathcal{H}_0$ as illustrated in case 2.1 of Fig. \ref{fig_case2}. 
\begin{figure}
\centering
\includegraphics[scale=1.0]{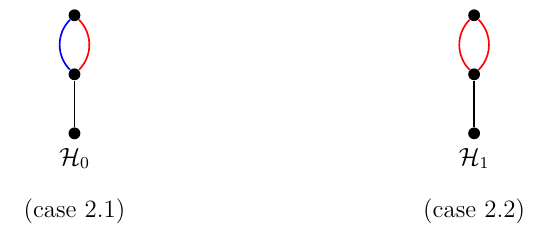}
        \caption{(case 2.1): The child of the reticulation vertex of the level-$1$ generator containing the root is attached to $\mathcal{H}_0$. A sequence of at least one network from $\mathcal{H}_0$ is attached to the red side. There exists exactly one network from $\mathcal{H}_1$ attached to the blue side which also may or may not contain level-$0$ networks attached to it. (case 2.2): The child of the reticulation vertex of the level-$1$ generator containing the root is attached to $\mathcal{H}_1$. A sequence of at least one networks from $\mathcal{H}_0$ is attached to each red side whose left-to-right order is irrelevant.}
\label{fig_case2}
\end{figure}

Exactly one level-$1$ network is attached to the blue side and there may or may not be level-$0$ network(s) attached to the blue side. Suppose that there are $k \geq 1$ network(s) that are attached to the blue side, exactly one of them is the level-$1$ network and the rest are level-$0$ networks. The level-$1$ network has $k$ options to be placed, therefore the networks attached to the blue side contribute $\sum_{k=1}^\infty kH_1(z)r(z)^{k-1}$. A sequence of at least one network of $\mathcal{H}_0$ is attached to the red side (otherwise this is a shortcut) and thus contributes $\frac{r(z)}{1-r(z)}$. Case 2.1 contributes $\sum_{k=1}^\infty kH_1(z)r(z)^{k-1}r(z)\frac{r(z)}{1-r(z)}=H_1(z)\frac{r(z)^2}{(1-r(z))^3}$.
\item A sequence of at least one network
from $\mathcal{H}_0$ is attached to each red side, whose left-to-right order is irrelevant and contributes $\frac{1}{2}H_1(z)\left(\frac{r(z)}{1-r(z)}\right)^2$ as seen in case 2.2 of Fig. \ref{fig_case2}.
\end{itemize}

Overall, Case 2 contributes $H_1(z)\frac{r(z)^2}{(1-r(z))^3}+\frac{1}{2}H_1(z)\left(\frac{r(z)}{1-r(z)}\right)^2$ to the exponential generating function
of $\mathcal{H}_2$.

\subsection{Case 3}

Any $N \in \mathcal{H}_2$ such that the root of $N$ belongs to a level-$2$ generator satisfies one of the four subcases:

\begin{itemize}
    \item Case 3.1: The child of the bottom reticulation vertex is the root of a network from $\mathcal{H}_0$ as seen in case 3.1 of Fig. \ref{fig_case_3a}. 
    
\begin{figure}[htb]
\centering
\includegraphics[scale=0.5]{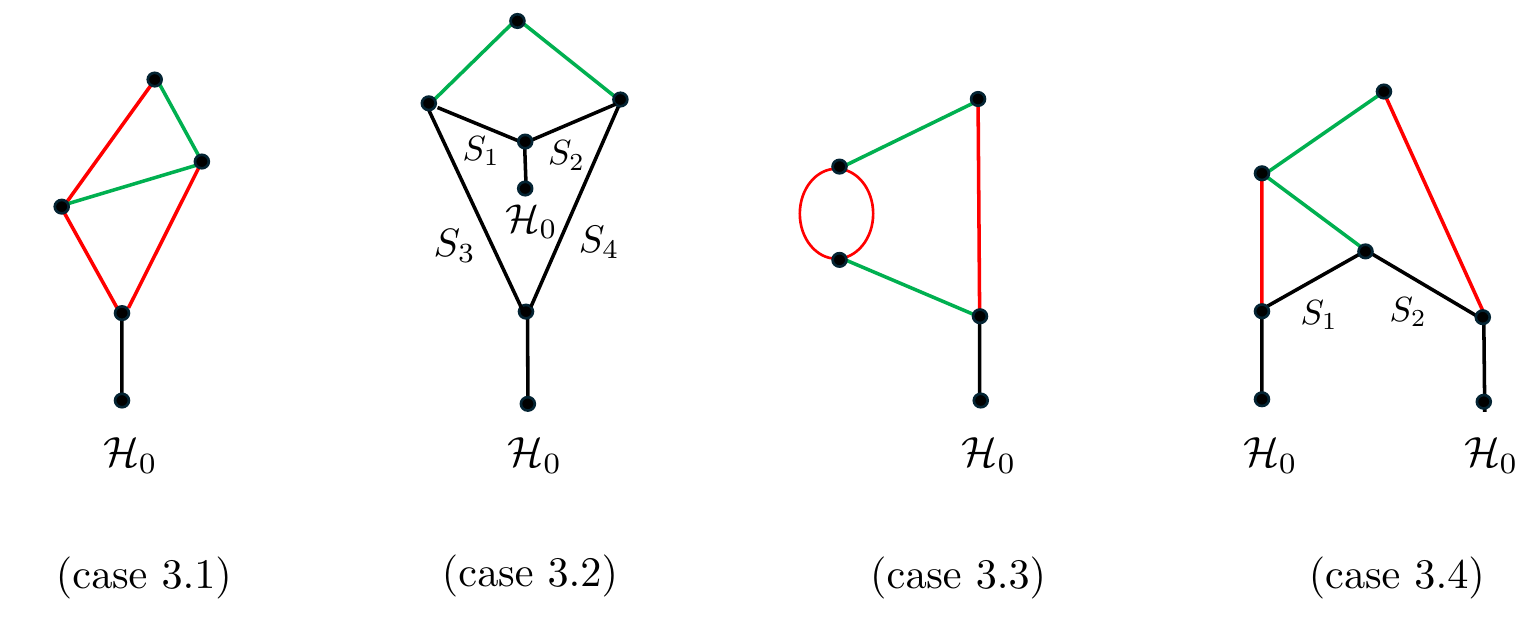}
\caption{(case 3.1): The child of the bottom reticulation vertex is attached to $\mathcal{H}_0$. The three red sides are non-empty while the two green sides can be empty. (case 3.2): Each child of the reticulation vertices is attached to $\mathcal{H}_0$. The two green sides can be empty. There are three subcases which depend on whether there are networks attached to $S_1,S_2,S_3,S_4$. The top-to-bottom and left-to-right orders are irrelevant. (case 3.3): The child of the bottom reticulation vertex is attached to $\mathcal{H}_0$. The three red sides are non-empty. The two green sides can be empty. The left-to-right order is irrelevant. (case 3.4): Each child of the reticulation vertices is attached to $\mathcal{H}_0$. The two red sides are non-empty. The two green sides can be empty. There are two subcases which depend on whether there are networks attached to $S_1$ and $S_2$.}
\label{fig_case_3a}
\end{figure}
        
All the three red sides cannot be empty, which means that at least one network from $\mathcal{H}_0$ is attached to them, otherwise the tree-child network and normal network conditions are violated. The two green sides can be empty, which means there may or may not be networks from $\mathcal{H}_0$ attached to them. This contributes $r(z)\left(\frac{r(z)}{1-r(z)}\right)^3\left(\frac{1}{1-r(z)}\right)^2$.
    
\item Case 3.2: The children of the reticulation vertices are the roots of networks from $\mathcal{H}_0$ as seen in case 3.2 of Fig. \ref{fig_case_3a} whose left-to-right and top-to-bottom orders are irrelevant. We consider the following situations:
\begin{itemize}
\item $S_1,S_2,S_3,S_4$ are all non-empty. The two green sides can be empty. This contributes 
$$\underbrace{\frac{1}{4}}_{\text{symmetry}}r(z)^2\left(\frac{r(z)}{1-r(z)}\right)^4\left(\frac{1}{1-r(z)}\right)^2.$$
\item Exactly one of $S_1,S_2,S_3,S_4$ is empty. The two green sides can be empty. This contributes
\begin{align*}
    \underbrace{4}_{\text{4 cases}}\cdot &\underbrace{\frac{1}{4}}_{\text{symmetry}} r(z)^2\left(\frac{r(z)}{1-r(z)}\right)^3\left(\frac{1}{1-r(z)}\right)^2\\
        &=r(z)^2\left(\frac{r(z)}{1-r(z)}\right)^3\left(\frac{1}{1-r(z)}\right)^2.
    \end{align*}
\item $S_1,S_2$ are empty, while $S_3,S_4$ are non-empty and vice versa. The two blue sides can be empty. This contributes
\begin{align*}
    \underbrace{2}_{\text{2 cases}}\cdot &\underbrace{\frac{1}{4}}_{\text{symmetry}} r(z)^2\left(\frac{r(z)}{1-r(z)}\right)^2\left(\frac{1}{1-r(z)}\right)^2\\
        &=\frac{1}{2}\left(\frac{r(z)}{1-r(z)}\right)^4.
\end{align*}
\end{itemize}
\end{itemize}

\begin{itemize}
    \item Case 3.3: The child of the bottom reticulation vertex is the root of networks from $\mathcal{H}_0$ as seen in case 3.3 of Fig. \ref{fig_case_3a}. All three red sides cannot be empty. The two green sides can be empty. The left-to-right order is irrelevant. This contributes
    $$\underbrace{\frac{1}{2}}_{\text{symmetry}} r(z)\left(\frac{r(z)}{1-r(z)}\right)^4\left(\frac{1}{1-r(z)}\right).$$
\end{itemize}

\begin{itemize}
    \item Case 3.4:  The children of the reticulation vertices are the roots of networks from $\mathcal{H}_0$ as seen in case 3.4 of Fig. \ref{fig_case_3a}. We consider two situations:
    \begin{itemize}
        \item Either $S_1$ or $S_2$ is empty. The two red sides cannot be empty otherwise they are shortcuts and the two green sides can be empty. This contributes 
        $$\underbrace{2}_{\text{2 cases}}  r(z)^2\left(\frac{r(z)}{1-r(z)}\right)^3\left(\frac{1}{1-r(z)}\right)^2.$$
        \item $S_1$ and $S_2$ are both non-empty. The two red sides cannot be empty. The two green sides can be empty. This contributes 
        $$r(z)^2\left(\frac{r(z)}{1-r(z)}\right)^4\left(\frac{1}{1-r(z)}\right)^2.$$
    \end{itemize}
\end{itemize}

In general, Case 3 contributes $\frac{1}{2}\frac{r(z)^4}{(1-r(z))^4}+\frac{r(z)^4}{(1-r(z))^5}+\frac{7}{2}\frac{r(z)^5}{(1-r(z))^5}+\frac{5}{4}\frac{r(z)^6}{(1-r(z))^6}$ to the exponential generating function of $\mathcal{H}_2$. 
\end{document}